\documentclass[fleqn,usenatbib]{mnras}

\usepackage{newtxtext,newtxmath}
\usepackage[T1]{fontenc}

\DeclareRobustCommand{\VAN}[3]{#2}
\let\VANthebibliography\thebibliography
\def\thebibliography{\DeclareRobustCommand{\VAN}[3]{##3}\VANthebibliography}

\usepackage{graphicx}	% Including figure files
\usepackage{amsmath}	% Advanced maths commands
\usepackage{bm}
\usepackage{mathtools}
\usepackage{soul} % Highlighting
\usepackage{float}
\usepackage{hhline} % for table formatting
\usepackage{booktabs} % for table formatting

\newcommand{\Lagr}{\mathcal{L}}

\title[Generative mass-mapping with fast UQ]{Generative modelling for mass-mapping with fast uncertainty quantification}

\author[Jessica J. Whitney et al.]{
Jessica J. Whitney,$^{1}$\textsuperscript{\thanks{E-mail: jessica.whitney.22@ucl.ac.uk}}
Tobías I. Liaudat,$^{2}$
Matthew A. Price,$^{1}$
Matthijs Mars,$^{1}$
Jason D. McEwen$^{1}$\textsuperscript{\thanks{E-mail: jason.mcewen@ucl.ac.uk}}
\\
$^{1}$ Mullard Space Science Laboratory (MSSL), University College London (UCL), Surrey RH5 6NT, UK\\
$^{2}$ IRFU, CEA, Université Paris-Saclay, F-91191 Gif-sur-Yvette, France
}

\date{Accepted 2025 August 13. Received 2025 July 31; in original form 2024 November 27}

\pubyear{2024}

\begin{document}
\label{firstpage}
\pagerange{\pageref{firstpage}--\pageref{lastpage}}
\maketitle

\begin{abstract}
	Understanding the nature of dark matter in the Universe is an important goal of modern cosmology. A key method for probing this distribution is via weak gravitational lensing mass-mapping---a challenging ill-posed inverse problem where one infers the convergence field from observed shear measurements. Upcoming stage IV surveys, such as those made by the Vera C.\ Rubin Observatory and Euclid satellite, will provide a greater quantity and precision of data for lensing analyses, necessitating high-fidelity mass-mapping methods that are computationally efficient and that also provide uncertainties for integration into downstream cosmological analyses.  In this work we introduce MMGAN, a novel mass-mapping method based on a regularised conditional generative adversarial network (GAN) framework, which generates approximate posterior samples of the convergence field given shear data. We adopt Wasserstein GANs to improve training stability and apply regularisation techniques to overcome mode collapse, issues that otherwise are particularly acute for conditional GANs. We train and validate our model on a mock COSMOS-style dataset before applying it to true COSMOS survey data. Our approach significantly outperforms the Kaiser-Squires technique and achieves similar reconstruction fidelity as alternative state-of-the-art deep learning approaches.  Notably, while alternative approaches for generating samples from a learned posterior are slow (e.g.\ requiring $\sim$10 GPU minutes per posterior sample), MMGAN can produce a high-quality convergence sample in less than a second. 
\end{abstract}


%=======================================================================================
\section{Introduction}\label{sec:introduction}
%=======================================================================================

The shape and magnitude of distant galaxies appear distorted under observation due to gravitational lensing, wherein the path of the photons emitted by these galaxies is bent by the gravitational field of intervening matter.
This distortion can be used to infer large-scale cosmological structure, in particular the distribution of said intervening matter---both visible matter and dark matter. Stage IV surveys, such as the Vera C. Rubin Observatory \citep{ivezic2019lsst} and Euclid \citep{laureijs2011euclid}, will provide
an abundance of new data for lensing analyses.

Weak lensing has two effects to first order: convergence, $\kappa$, and shear, $\gamma$. Mass-mapping is the process of approximating the convergence from the shear and is an ill-posed inverse problem due to instrumental and noise effects. Mass-maps are incredibly useful for calculating higher-order statistics---such as Minkowski functions and bispectrum \citep{munshi2017integrated}, peak count statistics \citep{liu2015cosmology,liu2015cosmological,martinet2018kids,harnois2021cosmic} and scattering transform statistics \citep{cheng2020new}---which can be compared against predictions for different cosmological models to constrain parameters and refine our understanding of the true underlying nature of the Universe. Given the advances in observational technology, we are now in the era of precision cosmology, where weak lensing measurements can provide unprecedented insights into the large-scale structure of the Universe \citep{mandelbaum2018weak}. To fully exploit the potential of these applications, mass-mapping methods must advance to provide precise and accurate reconstructions on par with the quality of the data now available to us. These methods should aim to retain small-scale structure and minimise the loss of information due to noise. Furthermore, they should provide well-characterised uncertainties, and be computationally feasible for large-scale inference to enable robust cosmological parameter estimation and model comparison.

The Kaiser-Squires technique \citep{kaiser1993mapping} is the seminal mass-mapping approach, based on direct inversion of the noisy shear field. It remains widely-used due to its speed and computationally efficiency. In an idealised setting it is equivalent to the maximum-likelihood estimator (MLE) due to the straightforward invertibility of the forward model. In practice, however, additional factors such as masking and instrumental effects break this invertibility, making mass-mapping an ill-posed problem \citep{price2021sparsecelest}.  Prior information is often injected to regularise ill-posed problems to recover effective solutions---an aspect not addressed by the Kaiser-Squires method. Instead, it is common to include a post-processing step in the Kaiser-Squires approach, typically Gaussian smoothing, to improve the signal-to-noise ratio of a reconstruction. However, this smoothing suppresses small-scale structures, potentially discarding valuable non-Gaussian information. While this may be an acceptable trade-off in certain applications (e.g.\ \citealt{kovacs2022view}), it is fundamentally limiting for higher-order statistical analyses and cosmological parameter estimation, where the small-scale structure is crucial. Wiener filtering \citep{wiener1949extrapolation} provides a \textit{maximum a posteriori} (MAP) estimate, assuming a Gaussian prior based on a fiducial cosmology for the convergence.  Like Gaussian-smoothed Kaiser-Squires maps, the Wiener filter also suppresses small-scale structures that are dominated by noise, resulting in a similar loss of information. Wavelet-based methods \citep{lanusse2016high,price2019sparse,price2020sparse,price2021sparse, price2021sparsehypoth,starck2021weak} aim to retain small-scale structure through the use of wavelet-based priors. These priors have led to increased performance, however, the lack of flexibility with hand-crafted priors has led to growing interest in deep learning methods, which can learn the priors from data itself.

Recent deep learning approaches for mass-mapping fall into two main categories. The first involves using deep learning for post-processing. For example, \citet{jeffrey2020deep} introduce a technique to post-process a reconstructed convergence field using a convolutional U-Net \citep{ronneberger2015u}. This approach is very fast but does not provide any uncertainty quantification. Conversely, \citet{shirasaki2021noise} use conditional generative adversarial networks (GANs) to learn noise maps which can then be used to denoise convergence maps.  Again, uncertainties are not provided. The second general category seeks to directly learn the posterior distribution of the data using deep learning. For example, \citet{remy2023probabilistic} introduce a technique that generates samples from a learned approximate posterior distribution, using data-driven priors learned by neural score estimation. These approximate posterior samples can then be used to construct a point-estimate reconstruction and to estimate uncertainties. However, it comes at the cost of being very slow, requiring 10 GPU minutes to generate a single independent approximate posterior sample. For both categories generative modelling techniques have proven highly effective due to the rich data-driven prior information that they are able to capture \citep{shirasaki2021noise,remy2023probabilistic}.  While all of these deep learning methods have provided promising results, particularly in recovering both large and small-scale structure in convergence maps, there is still work to be done to develop methods that provide high fidelity reconstructions, are computationally efficient, and also provide uncertainty estimates.  Fast generation of approximate posterior samples is necessary for integration into downstream cosmological parameter estimation and model comparison pipelines so that uncertainties in the mass-mapping process are captured.

To address these challenges we propose a novel mass-mapping method named MMGAN that is based on a regularised conditional GAN framework that generates approximate posterior samples of the mass-mapping inverse problem. Unlike \citet{shirasaki2021noise}, which uses conditional GANs to learn noise maps, our method attempts to learn the posterior distribution of the convergence field directly and sample from it. Furthermore, we adopt Wasserstein GANs \citep{arjovsky2017wasserstein} to improve training stability and apply regularisation techniques \citep{bendel2024regularized} to overcome mode collapse, issues that otherwise are particularly acute for conditional GANs. We show that MMGAN produces high-quality convergence samples, is highly computationally efficient, and provides accurate uncertainty estimates. We apply our model to both simulations and COSMOS survey data \citep{scoville2007cosmos}, and compare our results to the Kaiser-Squires method and \citet{remy2023probabilistic} to demonstrate its effectiveness.

The structure of this paper is as follows. In Section \ref{sec:background} we provide an overview of weak gravitational lensing and mass-mapping,
as well as an overview of GANs. In Section \ref{sec:methodology} we introduce our MMGAN approach. In Section \ref{sec:simulations_data} we describe how we constructed our training dataset, outline our approach to model training, and describe our validation and model selection methods. Then, in Section \ref{sec:results} we present our results, and in Section \ref{sec:conclusions} we discuss our conclusions.

%=======================================================================================
\section{Background}\label{sec:background}
%=======================================================================================

In this section we provide an overview of weak gravitational lensing and mass-mapping. For a more detailed review of weak lensing we refer the reader to \citet{bartelmann2001weak}. For a current review of the adoption of machine learning for astrophysics we refer the reader to \citet{lanusse2023dawes}. We also provide a brief overview of GANs, 
specifically conditional GANs. Further discussion on GANs can be found in related articles \citep{goodfellow2014generative, mirza2014conditional,goodfellow2020generative,creswell2018generative}.

\subsection{Weak Gravitational Lensing}\label{subsec:weak_lensing}

Distant sources emit photons which travel along space-time geodesics. In an empty universe, or one with uniformly distributed matter, these geodesics are simply straight lines, however this is not generally the case. 
Distributions of matter in the Universe, both visible and dark, induce local Newtonian potentials which result in perturbed geodesics, lensing the natural path of photons under gravity. When such perturbations are considered in aggregate our perception of distant objects is distorted. As these observable distortions are sensitive to all matter they are a natural cosmological probe for dark matter, dark energy, and the nature of gravity. Such distortions affect both the shape and apparent magnitude of the object, and the distant object is said to have been gravitationally lensed.

Suppose we consider photons which have an angular position on their source plane $\beta$, relative to the line-of-sight from observer through the primary lensing mass, greater than one Einstein radius $\omega_E$ from intervening matter, the lensing is said to be in the weak lensing regime. This ensures that the lensing effects are small, that the lensing equation 
\begin{equation}
	\beta = \omega - \omega_E \frac{\omega}{|\omega|^2}
	\qquad \text{where} \qquad
	\omega_E = \frac{4GM}{c^2} \frac{f_K(r-r^\prime)}{f_K(r)f_K(r^\prime)},
\end{equation}
is singular, and that distant objects cannot be multiply imaged. Here $G$ is the gravitational constant, $M$ is the lensing mass, $c$ is the speed of light, and $f_K$ denotes the angular diameter distance in the usual sense, which is dependent on the curvature $K$ of the Universe. The Universe has been observed to be essentially flat \citep{aghanim2020planck}. Consequently, it is often reasonable to approximate $K \approx 0 \Rightarrow f_K(r) \approx r$, where $r$ is the comoving distance. 

Consider now the local Newtonian potential $\Phi(r, \omega)$ induced by the matter distribution in the Universe, where $\omega = (\varphi, \vartheta)$ are spherical polar co-ordinates on the sky. Such physical potentials must necessarily satisfy Poisson's equation given by 
\begin{equation}\label{eq:poisson}
	\nabla^2 \Phi(r, \omega) = \dfrac{3\Omega_M H_0^2}{2 a(r)} \delta(r, \omega),
\end{equation}
where $\delta(r,\omega)$ denotes the fractional overdensity, $H_0$ is the Hubble constant, $a(r)$ is the scale-parameter, and $\Omega_M$ is the density of matter in the Universe. Integrating this potential along the line of sight produces the lensing potential 
\begin{equation}\label{eq:lensing_potential}
	\phi(r,\omega) = \dfrac{2}{c^2} \int_{0}^{r}dr^\prime \dfrac{f_K(r - r^\prime)}{f_K(r)f_K(r^\prime)}\Phi(r^\prime,\omega),
\end{equation}
which conceptually aggregates the effect of $\Phi(r,\omega)$ over $r$, \emph{i.e.} the potential of this collective mass to induce lensing effects. These equations are straightforwardly connected through Laplacian 
\begin{equation}
	\nabla^2\phi(r,\omega) = \dfrac{3\Omega_M H_0^2}{c^2} \int_{0}^{r}dr^\prime \dfrac{f_K(r - r^\prime)}{f_K(r)f_K(r^\prime)}\dfrac{\delta(r, \omega)}{a(r)}.
\end{equation}
At linear order such a lensing induces two distortions. Images are magnified by a convergence field $\kappa$ and their ellipticity is anisotropically stretched by a shear field $\gamma$. Both the convergence $\kappa$ and shear $\gamma$ fields can be related to the lensing potential $\phi$ by the following expressions (\textit{e.g.} \citealt{wallis2022mapping})
\begin{equation}
	\kappa(r,\omega) = \frac{1}{4}(\eth\bar{\eth} + \bar{\eth}\eth) \phi(r,\omega)
	\quad \text{and} \quad 
	\gamma(r,\omega) = \frac{1}{2}\eth\eth\phi(r,\omega),
\end{equation}
where $\eth$ denotes the spin-$s$ raising operator,
\begin{equation}
	\eth = -\sin^s\vartheta \big ( \partial_\vartheta + \frac{i\partial_\varphi}{\sin\theta} \big ) \sin^{-s}\vartheta
	\approx -(\partial_x + i\partial_y)
\end{equation}
and where $\bar{\eth}$ denotes the spin-$s$ lowering operator \citep{newman:1966, goldberg:1967},
\begin{equation}
	\bar{\eth} = -\sin^{-s}\vartheta \big ( \partial_\vartheta - \frac{i\partial_\varphi}{\sin\theta} \big ) \sin^{s}\vartheta
	\approx -(\partial_x - i \partial_y).
\end{equation}
In both cases the final inequality represents the appropriate approximation when one considers a field of view small enough to satisfy the flat-sky approximation, in which the sky may reasonably be parametrised through cartesian co-ordinates $x,y$ in the tangent plane. 

Substituting the flat-sky approximation of the $\eth$ and $\bar{\eth}$ into the expression for the shear and convergence one finds that 
\begin{equation}
	\kappa = \frac{1}{2}(\partial_{xx}+\partial_{yy}) \phi
	\quad \text{and} \quad 
	\gamma = \frac{1}{2}(\partial_{xx}-\partial_{yy}+2i\partial_{xy})\phi,
\end{equation}
where $\partial_{xx}$ is shorthand for $\partial_x\partial_x$, and where we have dropped the function arguments for notational brevity. Next we take the Fourier transform of these differential equations to find 
\begin{equation}
	\tilde{\kappa} = \frac{1}{2}(k_x^2+k_y^2) \tilde{\phi}
	\quad \text{and} \quad 
	\tilde{\gamma} = \frac{1}{2}(k_x^2-k_y^2+2ik_xk_y)\tilde{\phi},
\end{equation}
from which we can straightforwardly eliminate $\tilde{\phi}$ to find 
\begin{equation} \label{eq:lensing_inv_problem}
	\tilde{\gamma} = \frac{k_x^2 - k_y^2 + 2ik_xk_y}{k_x^2+k_y^2}\tilde{\kappa} = \mathbf{D}\tilde{\kappa} 
	\:\: \Rightarrow \: \:  \gamma = \mathbf{F}^{-1}\mathbf{DF}\kappa
\end{equation}
where $\mathbf{D}$ represents the Fourier mapping and $\mathbf{F}$ represents the Fourier transform. This expression is called the \emph{lensing forward model} and determines how one may map between convergence and shear fields. 

Ideally one would observe both the shear and convergence, each of which encodes subtly different and complementary cosmological information. Unfortunately, the brightness of a distant object is \emph{a priori} unknownable and therefore it is impossible to observe $\kappa$ directly. Importantly, the distribution of intrinsic galaxy ellipticities has zero mean $\langle \epsilon_s \rangle$ whilst the shear field has non-zero mean $\langle \gamma \rangle \not= 0$. Therefore by aggregating many ellipticity observations the net lensing effect may be distilled $\langle \epsilon_s + \gamma \rangle = \langle \epsilon_s \rangle + \langle \gamma \rangle \approx \langle \gamma \rangle$. 

The accuracy of this approximation is determined by the number of objects $N_\text{g}$ over which one averages. Making a central limit theorem argument the variance of the residual intrinsic shear component, colloquially referred to as the \emph{shape noise}, is approximately given by $var(\epsilon_s) \approx \sigma^2_\epsilon / N_\text{g}$, where $\sigma_\epsilon$ is the intrinsic ellipticity dispersion which is typically $\sim 0.37$. Given the typical magnitude of $\gamma \sim 0.05$ one need only average over $N_\text{g}\approx 30$ observations to recover a fair estimate of the shear. 

\subsection{Lensing Inverse Problem}\label{subsec:weak_lensing_inverse}

With observations of $\gamma$ to hand one may attempt to infer $\kappa$ by exploiting their Fourier space relationship. The most na\"ive algorithm by which $\kappa$ may be recovered given observations of $\gamma$ is by simply inverting this relation $\tilde{\kappa}_{\text{KS}} = \mathbf{D}^{-1} \tilde{\gamma}$, which is the original method developed by \citet{kaiser1993mapping}. As discussed in the introduction, in an idealised setting $\kappa_{\text{KS}}$ is equivalent to the maximum-likelihood estimator \citep{price2021sparsecelest}. However, in realistic scenarios noise contributions are overwhelmingly dominant and complex masking is present, thus the two estimators are by no means equivalent.

The Kaiser-Squires estimator is known for its computational efficiency and simplicity, however, it comes with several major drawbacks. First and foremost, it does not account for observational noise, which consequently propagates directly to the reconstruction. In order to use the mass-maps for cosmological inference, post-processing methods are conventionally used to improve their signal-to-noise ratio. This post-processing typically takes the form of Gaussian smoothing, which leads to the loss of non-Gaussian features in the convergence map. In particular, this results in a suppression of peaks in the reconstruction, and loss of small-scale structure---both of which are critical information for contemporary cosmology. Second, it does not provide any measure of the uncertainties associated with the reconstruction.

Since the Kaiser-Squires method was proposed there have been many other methods developed for mass-mapping, such as sparsity-based wavelet methods \citep{lanusse2016high, price2019sparse, price2020sparse, price2021sparsehypoth, starck2021weak} and deep learning architectures \citep{jeffrey2020deep,shirasaki2021noise, remy2023probabilistic}. Several of these methods have further been extended from the flat-sky to the sphere for wide-field mass-mapping \citep{wallis2022mapping,chang2018dark, price2021sparse}. Nevertheless, it is fair to say mass-mapping is by no means a solved problem. 

The original Kaiser-Squires method is quick and computationally efficient at the cost of loss of information. Deep learning techniques, which are data-driven, have shown promise in capturing the complexities of features in the data, but each approach has its own drawbacks. Post-processing learned denoising methods such as those by \citet{jeffrey2020deep} and \citet{shirasaki2021noise} are fast but lack principled uncertainty quantification. Neural score estimation methods such as those by \citet{remy2023probabilistic} provide uncertainty estimates but are slow at run-time. An additional question that warrants further study is the accuracy of machine learning methods when the only available training data is simulated, often for a single fiducial cosmology.

Overall, deep learning methods show great promise for ill-posed inverse problems such as mass-mapping, however, there is still need for deep learning methods which are fast, that produce high-fidelity reconstructions, and provide uncertainty quantification. We address this need with our proposed method, MMGAN.

\subsection{Generative Adversarial Networks}\label{subsec:gan}

We will now briefly review the GAN framework \citep{goodfellow2014generative}. GANs are comprised of two models: 
a generator, $G_\theta$, with parameters $\theta$ and a discriminator, $D_\phi$, with parameters $\phi$. During training, examples $x$ are drawn from the real data distribution $p_{\text{r}}(x)$, which is unknown to the model. The generator learns a distribution $p_{g}(x)$, from which it will output samples, $\hat{x}$. The aim of the generator is to match $p_{g}(x)$ as closely as possible to $p_{\text{r}}(x)$. The discriminator's role is to assess incoming data (which is a mix of real and generated samples), and decide whether it belongs to $p_{\text{r}}$ or $p_{g}$. In 
other words it aims to distinguish true samples from samples produced by the generator.

Both $G_\theta$ and $D_\phi$ are trained simultaneously to solve a two-player minimax game
\begin{equation}\label{eq:value_function}
	\begin{split}
		\underset{G_\theta}{\text{min}}\text{i}\underset{D_\phi}{\text{max}} V (G_\theta,D_\phi) = & \mathbb{E}_{x \sim p_{\text{r}}(x)}[\log D_\phi(x)] \\ 
		& + \mathbb{E}_{z \sim p_{z}(z)}[\log(1 - D_\phi(G_\theta(z)))].
	\end{split}
\end{equation}
for a value function $V$ \citep[see][]{goodfellow2014generative}, where $z$ is a latent variable drawn from a distribution $p_{z}(z) \sim \mathcal{N}(0, 1)$. Through training the generator will learn how to construct better samples, leading to a drop in performance of the discriminator. Consequently, this motivates the discriminator to once again learn how to differentiate the true data from the generated data, which will incentivize the generator to learn richer features of the data, in order to produce more convincing samples \citep{saxena2021generative}. It is this adversarial framework which allows GANs to produce such high-quality realisations after training.

GANs famously suffer from two main challenges during training:
\begin{enumerate}\label{list:training_difficulties}
	\item difficulty in converging;
	\item mode collapse.
\end{enumerate}

The generator and discriminator are both playing a minimax game, however, the game is a non-cooperative one; the optimal solution to such games is the Nash equilibrium. 
For GANs, this is equivalent to a discriminator which outputs a score $D_\phi = 0.5$ for all inputs, indicating it is unable to distinguish between real and generated samples.

In practice, it is difficult to reach Nash equilibrium, and the discriminator may become too good at distinguishing between real and generated samples. One may think 
this is a good thing that will lead to an improved rate of training, however, to those versed in game theory it will come as no surprise that it in fact leads to the opposite. 
This is because in non-cooperative games, an improvement for one player inherently causes a loss in performance for the other, as such a strong player will dominate the game. 
A perfect discriminator will output $D_\phi(x) = 1, \space \forall \, x \in p_{r}$ and $D_\phi(x) = 0,  \space \forall \, x \in p_g$.
When this happens, $\log(1 - D_\phi(G_\theta(z))) = 0$, and the generator's influence on the value function is lost. This can lead 
to the generator struggling, or failing entirely \citep{arjovsky2017principled}. 

Conversely, during training the generator may reach a local minimum in the learned probability space---this translates to a sample which is 
particularly good at fooling the discriminator, especially in relation to other nearby samples in the generator's distribution. 
In such cases, there is little inherent incentive for the generator to further explore the target probability distribution when called to generate samples. 
This is a problem known as \textit{mode collapse}. In the most extreme scenario this can lead to the generator producing 
the exact same output each time it is called---this is known as total mode collapse \citep{metz2016unrolled}. Mode collapse is also a problem when calculating uncertainties, as the loss of diversity in the generated samples leads to severe bias in the uncertainty estimates.

In short, regardless of whether the discriminator performs badly or well, the generator does not receive rich enough feedback to wholly represent the true data distribution.

\subsubsection{Wasserstein GANs}\label{subsubsec:wgan}

Wasserstein GANs were developed by \citet{arjovsky2017wasserstein} in order to tackle the difficulty in GAN training mentioned in the previous section. The overall idea is to 
use a new distance metric for the loss function, in order to provide a gradient which was more meaningful to the generator.

The Wasserstein-1 distance (also known as Earth Mover's distance) \citep{peyre2019computational} between two continuous distributions $p_r$ and $p_g$ may be expressed using the dual formation of the Wasserstein-1 distance,
\begin{equation}\label{eq:Wasserstein_distance}
	W_1(p_r, p_g) = \underset{\lVert f\rVert_L \leq 1}{\text{sup}} \mathbb{E}_{x \sim p_r} [f(x)] - \mathbb{E}_{\hat{x} \sim p_g}[f(x)],
\end{equation}
where $f$ is a 1-Lipschitz continuous function satisfying the constraint $\lVert f\rVert_L \leq 1$ and $\lVert \cdot \rVert_L$ represents the Lipschitz norm.

To provide an intuition for what this distance represents,
imagine two separate piles of dirt whose shapes may be described by distributions $p_r$ and $p_g$ respectively. The Wasserstein-1 distance between these two distributions is 
the minimum energy cost of moving the dirt in the second pile such that it is transformed from shape $p_g$ to shape $p_r$. This cost is proportional to the amount of units of dirt 
moved multiplied by how far each unit has been moved. A lower value of $W_1$ indicates a higher level of similarity between the two distributions \citep{rubner2000earth}. The Wasserstein-1 distance is a more generalisable metric than the Jensen-Shannon (JS) divergence used in traditional GANs, as rather than measuring the point-wise similarity between two distributions, it measures the cost of transporting one distribution to another. This means the Wasserstein-1 distance can be used to compare distributions with disjoint support, which is a problematic area for the JS divergence that often leads to vanishing gradients. These disjoint distributions are especially common in higher-dimensional data spaces, such as images, making the Wasserstein-1 distance a more suitable choice for GANs.

Unlike in traditional GANs, where the discriminator is a direct critic of the samples, in a Wasserstein GAN, the discriminator is trained to learn the optimal function $f$ to help estimate $W_1(p_r, p_{g})$. As the loss of the discriminator decreases, so does the Wasserstein-1 distance between the two distributions, implying that the generator's distribution $p_{g}$ is approaching the true distribution $p_r$ \citep{weng2019gan}.

One key difference between Wasserstein GANs and standard GANs is that where a perfect discriminator causes the generator in a standard GAN to fail, Wasserstein GANs actually rely on training the discriminator 
to convergence. For this reason, typically the gradients of the discriminator are updated more frequently than the generator, which also leads to improved stability 
during training.

%=======================================================================================
\section{Methodology}\label{sec:methodology}
%=======================================================================================

In this section we describe our methodology for building and training our mass-mapping GAN, coined MMGAN, a regularised conditional GAN. We begin 
by introducing conditional GANs, and highlight how they differ to standard GANs. We then introduce regularisation techniques utilised 
to overcome training issues, such as mode collapse and lack of convergence, traditionally faced by conditional GANs, before describing how these techniques are utilised to also provide uncertainty quantification.

\subsection{Conditional GANs}\label{subsec:conditional_gans}

Conditional GANs \citep{adler2018deep} differ from standard GANs in that they are conditioned on auxiliary input data, $y$, typically some kind of class or observational data, to which both the generator and discriminator have access. This additional information allows for greater control over the generated output, as the model is conditioned to provide targeted samples for a given input.

Consider the sets of data and observations $\mathcal{X}$ and $\mathcal{Y}$ respectively. The goal of the generator is to learn a generating function $G_\theta : \mathcal{Z} \times\mathcal{Y} \rightarrow \mathcal{X}$, where $\theta$ are the parameters of the generator, and $\mathcal{Z}$ is the set latent variables $z \sim p_z = \mathcal{N}(0, I)$. This function takes observations $y \in \mathcal{Y}$ as input, as well as some independently drawn $z$, and produces samples $\hat{x} = G_\theta (z,y)$. The role of the latent variable is to provide a source of randomness to the generator, such that even for fixed $y$, the generator can produce a variety of samples. Within a conditional GAN, the discriminator's function is of form $D_\phi : \mathcal{X} \times \mathcal{Y} \rightarrow [0, 1]$, with parameters $\phi$. The discriminator's role is still to determine whether a given samples is real or generated, however, it also has access to the observation $y$.

Each $\{x, y\}$ pair is unique, meaning there is only a single data instance, $x$, corresponding to data $y$. This can become a challenge when training conditional GANs and can lead to more acute mode collapse. Additionally, solutions 
to mode collapse in unconditional GANs \citep{schonfeld2020u,karras2020training,zhao2021improved} are often ill-suited to conditional GANs 
because of the presence of a conditioning variable. This means, that while Wasserstein GANs were sufficient to effectively solve both issues of unstable training and mode collapse for standard GANs, in conditional GANs these challenges require distinct solutions.

\subsection{Conditional Wasserstein GANs}\label{subsec:conditional_wgan}

One can still use the notion of Wasserstein-1 distance, and adapt it for conditional data as follows
\begin{equation}
	W_1(p_r(\cdot | y), p_g(\cdot | y)) = \underset{D_\phi\in L_1}{\text{sup}}  \{ \mathbb{E}_{x \sim p_r} \{ D_\phi(x|y)\} -\mathbb{E}_{\hat{x} \sim p_g}\{ D_\phi(\hat{x}|y)\}\}, 
\end{equation}
where $x$ is a true sample, with observation $y$, $L_1$ is the set of 1-Lipschitz continuous functions, and $\hat{x}$ is a generated sample for that observation. Through using the discriminator to estimate this Wasserstein-1 distance, the resulting conditional GAN still benefits from increased training stability, while avoiding the vanishing gradient problem.

\subsection{Regularised conditional GANs}\label{subsec:regularised_cgan}

Regularised conditional GANs \citep{bendel2024regularized} are a recent development, designed to overcome mode collapse in conditional GANs---which as previously mentioned is a more acute problem than with traditional GANs, and also harder to solve due to the one-to-one ${x, y}$ pairing of data. Within this framework, the generator aims to solve the following minimisation problem
\begin{equation}
    \text{arg min}_\theta \{ \beta_{\text{adv}}\Lagr_{\text{adv}}(\theta, \phi) + \Lagr_{1, \text{SD}, N_{\text{train}}}(\theta, \beta_{\text{SD}})\},
\end{equation}
where $N_{\text{train}} \geq 2$ represents the number of samples made by the generator, and $\beta_{\text{adv}}$ and $\beta_{\text{SD}}$ are hyperparameters which control the relative importance of the adversarial loss term $\Lagr_{\text{adv}}$ and the regulariser $\Lagr_{1, \text{SD}, N_{\text{train}}}$, respectively. They themselves are defined as
\begin{equation}
	\Lagr_{\text{adv}}(\theta, \phi) := \mathbb{E}_{x,z,y} \{ D_\phi (x|y) - D_\phi (G_\theta (z|y)|y)\},
\end{equation}
and
\begin{equation}
    \Lagr_{1, \text{SD}, N_{\text{train}}}(\theta, \beta_{\text{SD}}) := \Lagr_{1, N_{\text{train}}}(\theta) - \beta_{\text{SD}}\Lagr_{\text{SD}, N_{\text{train}}}(\theta).
	\label{eq:regulariser}
\end{equation}

As can be seen in Equation \eqref{eq:regulariser}, the regulariser is a combination of two loss functions: 
the first being the $N_{\text{train}}$-sample supervised $\ell_1$ loss; and the second being the standard 
deviation reward. These losses are defined by
\begin{equation}
    \Lagr_{1, N_{\text{train}}}(\theta) := \mathbb{E}_{x, z_1, ...., z_N, y}\{ \lVert x - \hat{x}_{(N_{\text{train}})}\rVert_1\},
\end{equation}
and 
\begin{multline}
	\Lagr_{\text{SD}, N_{\text{train}}}(\theta) := \sqrt{\dfrac{\pi}{2 N_{\text{train}}(N_{\text{train}} - 1)}} \\
	\times \sum_{i=1}^{N_{\text{train}}}\mathbb{E}_{z_1,...,z_N,y} \{ \lVert\hat{x}_{i} - \hat{x}_{(N_{\text{train}})}\rVert_1\}
\end{multline}
where $\{ \hat{x}_{i}\}$ are the generated samples and 
$\hat{x}_{(N)} := 1/N \,\sum_{i=1}^{N} \hat{x}_{i}$ is the $N$-sample average. By including the standard deviation within the reward function, the model is encouraged to produce samples with some diversity, which helps to avoid mode collapse.

The choice of $\ell_1$-loss and standard deviation reward is not an arbitrary one. It can be shown that in the case where the generated samples $\hat{x}_i$ and the true samples $x$ are both independent Gaussian distributions conditioned on $y$, the mean and covariance of the generated samples will match that of the true distribution \citep[\S Prop. 3.1]{bendel2024regularized}. That is to say,
\begin{equation}\label{eq:prop_3.1_expectation}
	\mathbb{E}_{z_i \sim p_z} \{ \hat{x}_i (\theta^*) | y\} = \mathbb{E}_{x \sim p_r} \{ x | y\} = \hat{x}_{\text{MMSE}}
\end{equation}
where $\hat{x}_{\text{MMSE}}$ is the minimum mean squared error (MMSE) estimate of the true posterior, and
\begin{equation}\label{eq:prop_3.1_covariance}
	\text{Cov}_{z_i \sim p_z} \{ \hat{x}_i (\theta^*) | y\} = \text{Cov}_{x \sim p_r} \{ x | y\},
\end{equation}
where $\theta^* = \text{arg min}_\theta \Lagr_{1, \text{SD}, N_\text{train}}(\theta, \beta_\text{SD}^\mathcal{N})$ with $\beta_\text{SD}^\mathcal{N} := \sqrt{2 / (\pi N_\text{train}(N_\text{train}+1))}$ being the optimal parameters for the generator \citep{bendel2024regularized}.

In practice, the assumptions required for this proposition do not necessarily hold, therefore automatic tuning of the hyperparameter $\beta_{\text{SD}}$ is considered, which controls the desired standard deviation between generated approximate posterior samples. Some level of deviation between samples is necessary to avoid mode collapse, however, too much deviation among samples can hinder the model's ability to learn the true distribution of the data.

In order to constrain the allowed variance of generated samples, the model auto-tunes $\beta_{\text{SD}}$ during training. The method utilises an observation made by \citet[\S Prop. 3.3]{bendel2024regularized} that when $\hat{x_i} \sim p_r(\cdot | y)$ are independent samples of the true posterior, then the ratio between the $\ell_2$ error of a single sample and the $N$-average sample is given by
\begin{equation} 
	\dfrac{\varepsilon_1}{\varepsilon_N} = \dfrac{2N}{N+1},
\end{equation}
where $\varepsilon_1$ and $\varepsilon_N$ are approximated as follows
\begin{equation}
	\hat{\varepsilon}_1 = \dfrac{1}{N_\text{val}} \sum_{i=1}^{N_\text{val}} \lVert x_i - \hat{x}_1\rVert^2_2,
\end{equation}
and
\begin{equation}
	\hat{\varepsilon}_N = \dfrac{1}{N_\text{val}} \sum_{i=1}^{N_\text{val}} \lVert x_i - \sum_{j=1}^{V}\hat{x_j}\rVert^2_2,
\end{equation}
for some validation set $\{(x_v, y_v)\}^{N_\text{val}}_{v=1}$. This ratio is calculated during each training epoch $\tau$. Then, $\beta_{\text{SD}}$ is updated using gradient descent according to the following equation
\begin{equation}
	\beta_{\text{SD}, \tau+1} = \beta_{\text{SD}, \tau}  - \mu_{\text{SD}} \left( \log_{10} \left[ \dfrac{\hat{\varepsilon}_{1,\tau}}{\hat{\varepsilon}_{N_\text{val}, \tau}} \right] - \log_{10}\left[ \dfrac{2N_\text{val}}{N_\text{val}+1}\right]\right)\beta_{\text{SD}, \tau=0}
	\label{eq:beta_SD_update}
\end{equation}
for a learning rate $\mu_{\text{SD}} > 0$. For the full details of the above proposition we refer the reader to \citet{bendel2024regularized}.

\subsection{MMGAN}\label{subsec:model_architecture}

With all the necessary components described, we now introduce our model architecture. Our regularised conditional GAN, MMGAN, follows the same general structure as \citet{bendel2024regularized}, with some key changes to tailor the model to mass-mapping. The goal of our model is to produce approximate posterior samples of the convergence given a shear map. 

Our generator is based on a U-Net architecture \citep{ronneberger2015u}. There are 6 input channels: the shear map; a Kaiser-Squires reconstruction (made on-the-fly from the shear map) with no added smoothing; and a random noise vector $z \sim N(0,I)$. Each of these inputs includes two channels, one for the real component of the input and another for the imaginary component. We trialled models both with and without the Kaiser-Squires map as an additional input channel, and found the addition of it led to improved performance, with negligible increase in computational cost. In particular we observed that the shear alone is good for capturing the uncertainties within a reconstruction, however, the addition of the Kaiser-Squires map leads to better quality reconstructions. Although we used Kaiser-Squires here, for its simplicity, it is worth noting that any fast, approximate reconstruction method could be used in its place. Other mass-mapping methods \citep{jeffrey2020deep, shirasaki2021noise} also take an approximate reconstruction as input, although they typically do not condition on the observed shear field as we do here, (\textit{e.g.} \citealt{jeffrey2020deep} post-process a Wiener filter reconstruction).

Our network architecture consists of 4 downsampling blocks, starting with 128 initial channels. Rather than traditional pooling methods, we downsample through convolutional blocks. Each block consists of a convolutional layer with a kernel of size $3\times3$ and padding of 1, followed by batch normalisation and a Parametric ReLU (PReLU) activation function. At this point, we include a residual block, which consists of two convolutional layers followed by batch normalisation and a PReLU activation function. This block is our skip connection. Then, we take a final convolutional layer with a kernel of size $3\times3$, padding of 1, and stride of 2, which will act as our downsampling step. The number of channels doubles at each downsampling block, from 128 to 256, 512, and finally 1024. 

\begin{figure*}
	\centering
	\includegraphics[width=\textwidth]{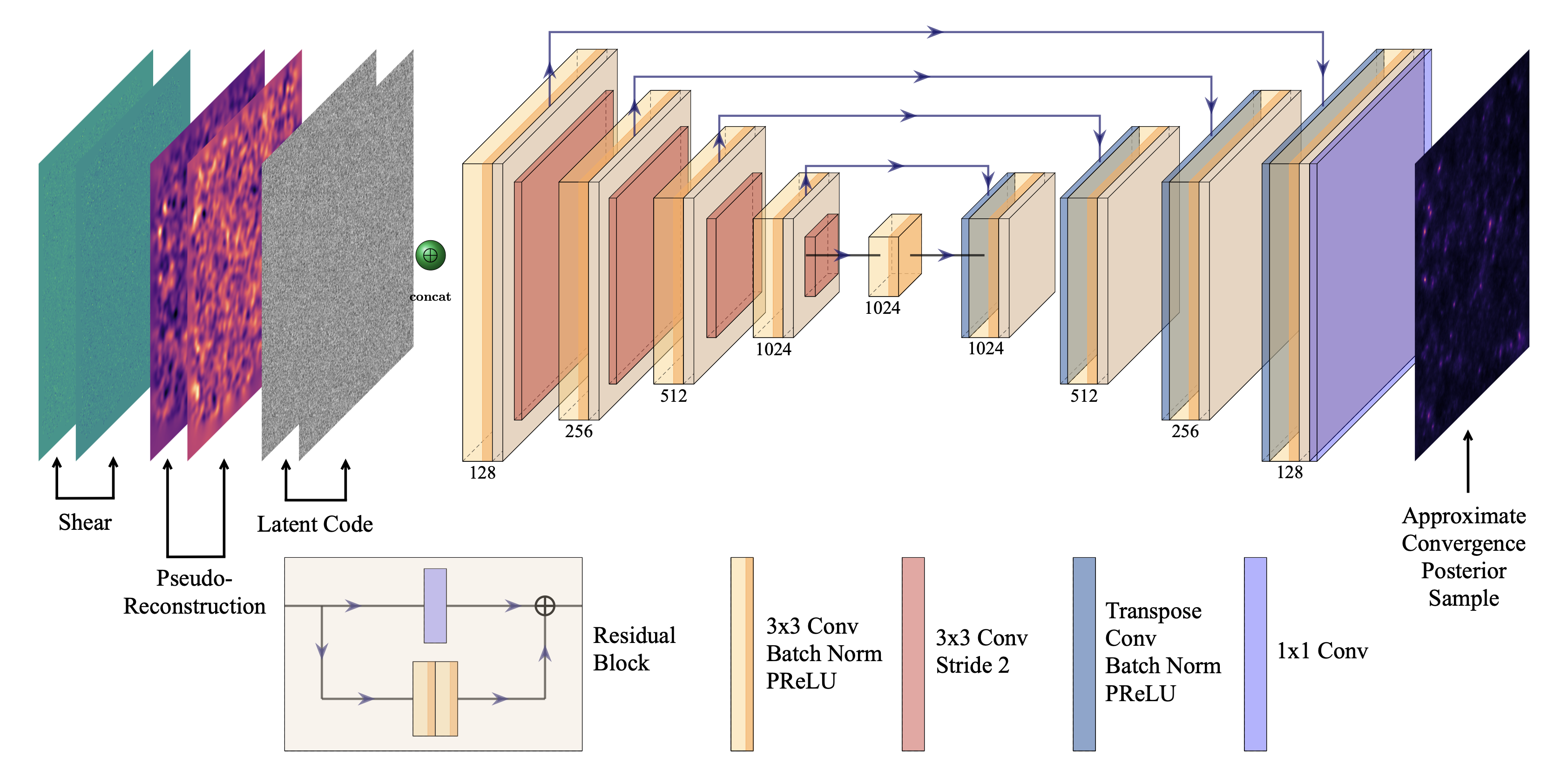}
	\caption{Illustration of the architecture of the MMGAN generator. The shear map, comprising real and imaginary components, is used to produce a pseudo-reconstruction, which is similarly decomposed into its real and imaginary parts. These components are concatenated with a two-channel random latent vector $z$, and subsequently passed through our U-Net generator, which outputs a single sample of the convergence from the learned posterior distribution. The numbers below each block indicate the number of channels in each layer. Additionally, the colour of the blocks indicate the series of operations applied, as dictated by the legend. The residual block has been illustrated in greater detail below the main generator architecture. It consists of a $3\times3$ convolution, followed by batch normalisation, then parametric ReLU. This is done twice, and then the output of this is added to a $1\times1$ convolution of the original input.}\label{fig:generator}
\end{figure*}

In the bottleneck of the U-Net, we include a single residual block before moving to our upsampling blocks. For upsampling, we rely on transpose convolutions. The number of channels halves at each upsampling block; as such, the number of output channels is also 128. Each upsampling layer begins with a transpose convolutional layer with kernel size $3\times3$, padding size 1 and stride 2, which acts as our upsampling mechanism and is followed by batch normalisation and a PReLU activation function. We then concatenate the output with the corresponding skip connection and again apply a convolutional layer with kernel size $3\times3$ and padding size 1, followed by batch normalisation, PReLU activation, and a residual block. As we move through the upsampling blocks, the number of channels halves, meaning after 4 layers, the number of channels is once again 128. Finally, after upsampling, we apply two convolutions with $1\times1$ kernels. The output of the generator is a single approximate sample of the convergence field. An illustration of the generator can be found in Figure \ref{fig:generator}.

Our discriminator is a standard CNN classifier, taking two inputs $x$ and $y$, with one initial convolutional layer followed by 6 convolutional downsampling layers and one final fully-connected layer. In the initial layer, we use convolutions with kernels of size $3\times3$ with 1 padding and a leaky ReLU with a negative slope of 0.01. In each downsampling layer, we reduce spatial resolution with average pooling, using $2\times2$ kernels with a stride of 2, then a convolutional layer with a $3\times3$ kernel with padding of 1, instance normalisation, and a leaky ReLU with a negative-slope of 0.2. The final output of the discriminator is the estimated Wasserstein score for the convergence map.

\subsubsection{Point Image Estimate}\label{subsubsec:point_image_estimate}

In order to create a final convergence map reconstruction we need to select a suitable point estimate. It is natural to use the posterior mean, which is also the MMSE, especially given Equation \eqref{eq:prop_3.1_expectation}, where it is shown that under certain assumptions a link can be drawn between the MMSE and the true posterior. Therefore, to build the final convergence map, the shear map is passed through the generator many times. Each time the generator is called it produces a new approximate posterior sample. The empirical posterior mean is used as the MMGAN reconstruction, which is obtained by doing an average of $N$ approximate posterior samples. 

\subsubsection{Uncertainty Quantification}\label{subsubsec:uncertainty_quantification}

The convergence map reconstruction is the average of $N$ approximate posterior samples. For uncertainty quantification, 
we calculate the pixel-wise standard deviation of the samples, in order to build an uncertainty map.
Based on the proposition outlined in \citet{bendel2024regularized}, the standard deviation of the approximate posterior samples matches that of the true posterior, under certain assumptions. Therefore, features that consistently appear across the generated samples are more likely to be true features of the data, as compared to features which appear in one or two samples. For features that appear in the majority of samples, 
the standard deviation for those pixels will be low. Conversely, in areas where the model is less 
certain about present features, the generated samples will be more diverse, meaning the standard 
deviation of that region will be higher. In this way, by looking at the standard deviation map,
one can infer the model's confidence in the reconstruction.

%=======================================================================================
\section{Simulations, Training, and Validation}\label{sec:simulations_data}
%=======================================================================================

This section details the simulations and mock dataset used to train our model. We first discuss the $\kappa$TNG simulations, a collection of convergence maps based on the IllustrisTNG simulations, before moving on to describe how we used this weak lensing map suite to build a mock catalog of 10,000 convergence maps in the style of the COSMOS survey. This catalog was then used to train, validate, and test our model.

\subsection{KappaTNG Simulations}\label{subsec:kappatng}

The $\kappa$TNG simulations are a suite of $10,000$ mock weak lensing maps \citep{osato2021kappatng}, 
based on the IllustrisTNG hydrodynamical simulations \citep{springel2018first}. All 
simulations assume the flat $\Lambda$CDM cosmology as in Planck 2015 \citep{ade2016planck}, 
with $H_0 = 67.74 \: \text{kms}^{-1}\text{Mpc}^{-1}$, baryonic density $\Omega_b = 0.0486$, 
matter density $\Omega_m = 0.3089$, and spectral index of scalar perturbations $n_s = 0.9667$.

The maps were generated by creating light cones with an opening angle of $5 \times 5 \: \text{deg}^2$, 
from the IllustrisTNG simulations, made by stacking TNG snapshots along the line of sight. 
The mock weak lensing maps were then created by tracing the light cones from $z=0$ up to the target redshift, 
$z_s \in [0.00, 2.57]$. To create the full suite, a large number of random flips, rotations, and translations 
were applied to the IllustrisTNG snapshots. The subsequent maps were shown to be statistically independent \citep{osato2021kappatng}. 
Each map is of size $1024 \times 1024$ pixels, with a resolution 
of $0.29 \: \text{arcmin/pixel}$. 

\subsection{COSMOS Data}\label{sec:cosmos_data}

In the following analysis we make use of data from the COSMOS survey \citep{scoville2007cosmos}. The 
COSMOS field is a $1.64 \: \text{deg}^2$ field on the sky, images using the advanced camera for surveys (ACS). 
Throughout this work, we use the \citet{schrabback2010evidence} shape catalog, which is a 
catalog with two subsets: a bright catalog with $i^+ < 25$, and a faint catalog with $i^+ > 25$. Galaxy samples in the bright catalog can be cross-matched with the COSMOS-30 catalog \citep{ilbert2008cosmos}, 
providing individual photometric redshifts. This is not available for the faint catalog.

In our analysis we use the full catalog, including both the bright and faint samples. We cut bright galaxies with 
$z_{\text{phot}} < 0.6$ and $i^+ > 24$, as there are indications these may in fact be galaxies at 
high redshifts \citep{schrabback2010evidence}; see also \citet{remy2023probabilistic} for further discussion on this. 
After applying these cuts, the total number of galaxies is $417,117$.

\subsection{Mock COSMOS Dataset}\label{subsec:mock_maps}

In order to create mock COSMOS maps we utilised both the $\kappa$TNG simulations, and the 
\citet{schrabback2010evidence} shape catalog. As mentioned, this shape catalog is divided into a 
bright and faint catalog, which we combined into a full catalog. We discarded galaxies with photometric 
redshifts $z_{\text{phot}} < 0.6$ for reasons specified in Section \ref{sec:cosmos_data}. Then we calculated 
the redshift distribution, $p(z)$, of the galaxies in the full catalog.

Convergence maps for sources with a redshift distribution $p(z)$ calculated by
\begin{equation}\label{eq:kappa_tot}
	\kappa^{\text{tot}} = \sum_{i=i_{\text{min}}}^{i_{\text{max}}}w_i \kappa_i,
\end{equation}
where
\begin{equation}\label{eq:kappa_tot_weights}
	w_i = \int_{z_i - \Delta z_i/2}^{z_i + \Delta z_i/2} \text{d}z p(z),
\end{equation}
where $\kappa^{\text{tot}}$ is the convergence map for sources with a redshift distribution $p(z)$, $\kappa_i$ 
is the convergence map for sources at redshift $z_i$, $\Delta z_i$ is the width of the $i_{\text{th}}$-redshift bin, 
and $i_{\text{min}}$ and $i_{\text{max}}$ are the minimum and maximum redshifts of source galaxies considered, respectively 
\citep{makiya2021ray}.

The $\kappa$TNG maps are sliced at discrete redshifts between $z \in [0, 2.568]$, leading to $40$ evenly-spaced source planes. For our mock COSMOS maps we required a redshift up to $z=5$, therefore we chose redshift values $z_i$ with spacing equal to the $\kappa$TNG slices. This resulted in $80$ redshift values $z_i$, leading to $79$ redshift bins, centered on the redshift values of the $\kappa$TNG maps. 
Note that the bin size is halved for the first and last bins. For $z > 2.568$ we follow \citet{remy2023probabilistic} and reused the highest redshift slice convergence map ($z=2.568$), while calculating a new weight for each bin.

Finally, we created a mask to represent the COSMOS survey area. To do this we binned the COSMOS shape catalog into maps of the shear components, and calculated the number of galaxies per pixel. 
We created a binary mask for empty pixels.

In summary, we created $10,000$ convergence maps of size $300\times300$ pixels. In order to convert these to mock shear maps, we used the forward model described by Equation \eqref{eq:lensing_inv_problem}. We added spatially varying noise to the shear, treating the real and imaginary components separately by calculating the standard deviation of the $\gamma_1$ and $\gamma_2$ estimates in the COSMOS shape catalog respectively. We then simulated the noise by multiplying the two standard deviations by a random normal distribution, and adding to the real and imaginary components of the clean mock COSMOS shear maps. We note that these maps do not account for instrumental effects such as photometric uncertainty, intrinsic galaxy alignment, or baryonic feedback effects. However, if these effects were incorporated into the training data, the model could, in principle, learn to account for them.

\subsection{Training}\label{subsec:training}

During each training epoch, for a batch size and $N_{\text{train}}=2$, we generated mock shear maps on-the-fly from our convergence maps (see Section \ref{subsec:mock_maps}). Each shear map was paired with two latent vectors, corresponding to the real and imaginary components, which were then input to the generator. The generator optimised the following loss function
\begin{equation}
	\Lagr_{G_{\theta}} := \beta_{\text{adv}}\Lagr_{\text{adv}}(\theta, \phi) + \Lagr_{1, N_{\text{train}}}(\theta) - \beta_{\text{SD}}\Lagr_{\text{SD}, N_{\text{train}}}(\theta),
\end{equation}
where $\beta_{\text{adv}}$ was initially set to $10^{-2}$ for the first 5 epochs, then decreased to $10^{-4}$ until epoch 23, and finally to $10^{-5}$ for the remainder of training. The value of $\beta_{\text{SD}}$ was updated according as described in Section \ref{subsec:regularised_cgan} using $N_{\text{val}}=8$. Following this, the discriminator performed an optimisation step on its own loss
\begin{equation}
	\Lagr_{D_{\phi}} := -\Lagr_{\text{adv}}(\theta, \phi) + \alpha_1 \Lagr_{\text{grad}}(\phi) + \alpha_2 \Lagr_{\text{drift}}(\phi),
\end{equation}
where $\Lagr_{\text{grad}}$ is a gradient penalty used to encourage that $D_{\phi} \in L_1$ \citep{gulrajani2017}, with $\alpha_1=10$ the gradient penalty weight. We follow \citet{karras2018progressive} and add the term $\Lagr_{\text{drift}}$, which penalises the discriminator's output from drifting too far away from zero, as it can make the training unstable. More precisely, the drift penalty is defined as $\Lagr_{\text{drift}}(\phi) := \mathbb{E}_{x,y}\{ D_{\phi}(x|y)^2 \}$. Following \citet{adler2018deep}, we use a small drift penalty weight of $\alpha_2=0.001$. We used the Adam optimiser \citep{kingma2014adam} with a learning rate of $10^{-3}$, $\beta_1=0$, and $\beta_2=0.99$. Our model was trained across 4 Nvidia A-100 GPUs, and took approximately 6.5 hours to train for 100 epochs.

\subsection{Model Validation}\label{sec:model_selection}

To evaluate the performance of the trained model we validated it on a subset of our mock COSMOS dataset previously unseen by the model.

For validation we looked at the peak signal-to-noise ratio (PSNR), 
\begin{equation}
	\text{PSNR} = 10 \log_{10}\left(\dfrac{\text{MAX}_I^2}{\text{MSE}}\right),
\end{equation}
where $\text{MAX}_I$ is the maximum possible pixel value (which we set to 1), and MSE is the mean squared error between the truth and the reconstruction. We calculated the PSNR of a single posterior sample as well as the PSNR of a reconstruction made from the average of $N=32$ samples (for discussion on our choice of $N$ see Section \ref{subsubsec:reconstructions_sim}). 

This procedure is repeated across a number of different input maps. We then collate the data and calculate the difference between the single PSNR and reconstruction PSNR. We define a tolerance, and if the magnitude of the difference is larger than that tolerance we remove the epoch from the set. We took this approach to ensure that any epoch we considered as our final model was not one prone to over-variance in the generated samples.

With the remaining epochs, we calculated a range of metrics across a set of mock maps and reconstructions. These metrics were the PSNR, RMSE, and Pearson correlation coefficient,
\begin{equation}
	r = \dfrac{\sum\limits_i(x_i - \hat{x})(p_i - \hat{p})}{\sqrt{\sum\limits_i(x_i - \hat{x})^2}\sqrt{\sum\limits_i(p_i - \hat{p})^2}},
\end{equation}
where $x_i$ and $p_i$ are the truth and the reconstruction, respectively, and $\hat{x}, \hat{p}$ denote their respective means. We selected the epoch which performed best across all metrics as our final model.

%=======================================================================================
\section{Results}\label{sec:results}
%=======================================================================================

In this section we present both the performance of our model on the mock COSMOS test simulations, followed by its application to the true COSMOS field data. We show some example reconstructions and discuss how the quality of the reconstruction changes with the number of approximate posterior samples used to create it. We also compare our results to the Kaiser-Squires method, and in the case of our full COSMOS reconstruction we show it alongside the \citet{remy2023probabilistic} reconstruction. In addition, we show how the standard deviation map of the approximate posterior samples can be used to quantify the uncertainty in the reconstruction.

To assess reconstruction quality, we compared to the pixel-wise absolute error $= |\hat{x}_{(N)} - x|$ between the reconstruction and the ground truth. We also used the absolute error to asses the usefulness of using the standard deviation of the generated approximate posterior samples as a measure of uncertainty.

\subsection{Simulations}\label{sec:sim_results}

In this subsection we present the results of our model applied to the mock COSMOS simulations. We begin by motivating the choice of $N=32$ for the number of approximate posterior samples used to create a reconstruction. We then show some example reconstructions and approximate posterior samples. We directly compare our results to the Kaiser-Squires method, as well as qualitatively compare to other state-of-the-art methods. Finally, we assess the quality of our uncertainty quantification and calculate coverage probabilities for our reconstructions.

\subsubsection{Reconstructions of Simulations}\label{subsubsec:reconstructions_sim}

\begin{figure}
	\centering
	\includegraphics[width=0.48\textwidth]{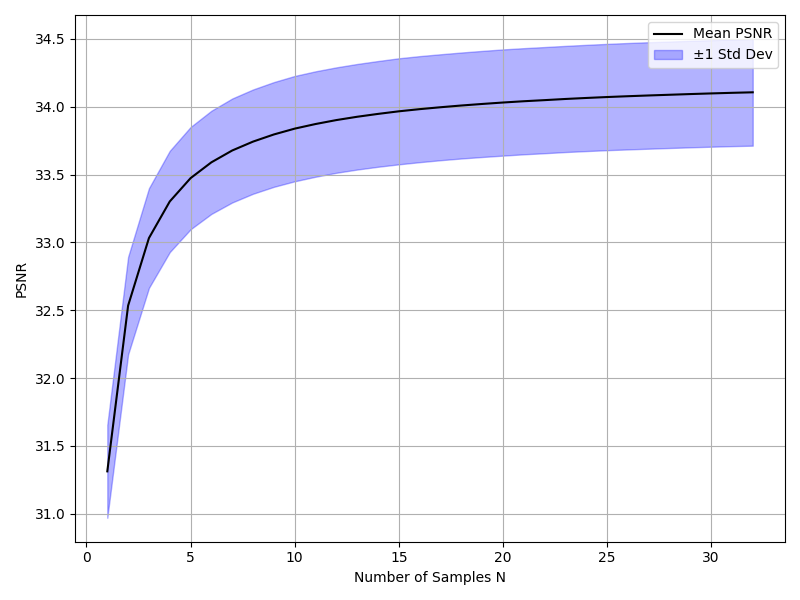}
	\includegraphics[width=0.48\textwidth, trim={0, 1cm, 0, 0}]{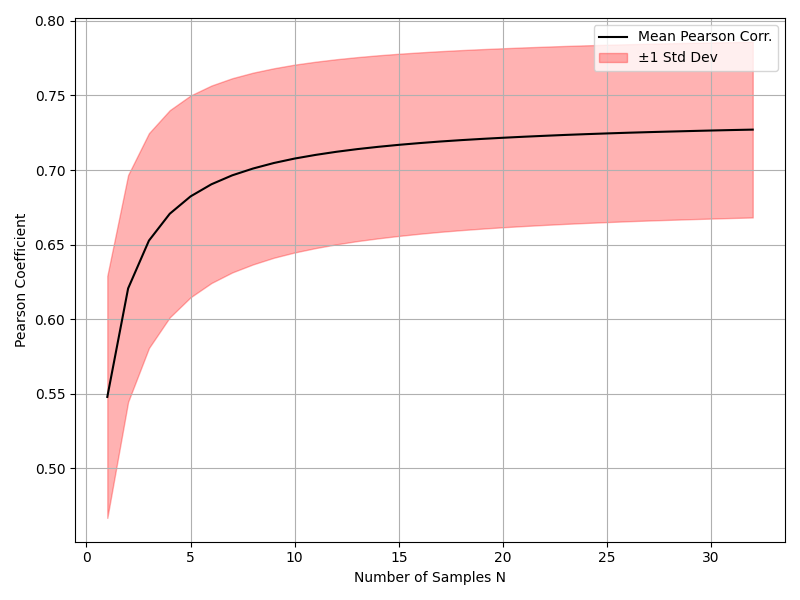}
	\caption{PSNR and Pearson correlation coefficient values of MMGAN reconstruction dependant on the number of approximate posterior samples used to create that reconstruction, which is given by the mean of the approximate posterior samples. The curve flattens out for both metrics, indicating there is little need to consider $N > 32$.}
	\label{fig:psnr_vs_p}
\end{figure}

When building a reconstruction it is important to choose an appropriate number of generated samples from which to calculate the mean. We calculated both the PSNR and the Pearson correlation coefficient for reconstructions made with different numbers, $N$, of samples. For both metrics, a higher value indicates a better reconstruction. Figure~\ref{fig:psnr_vs_p} illustrates how these metrics change with the number of samples used to create the reconstruction.
As can be seen, there is a large increase in quality between $N=1$ and $N=4$, however, the curves quickly flatten out for larger $N$ indicating that from a quality perspective there is no need to choose an excessively high value for $N$. We choose $N=32$ henceforth for reconstructions.

Table \ref{tab:results} further explores how the reconstruction changes as the number of generated samples used changes. It shows the PSNR and structural similarity index measure (SSIM) \citep{wang2004image} on reconstructions with different values of $N$, calculated during model validation. The SSIM is computed as 
\begin{equation}
	\text{SSIM}(x, y) = \dfrac{(2\mu_x\mu_y + C_1)(2\sigma_{xy} + C_2)}{(\mu_x^2 + \mu_y^2 + C_1)(\sigma_x^2 + \sigma_y^2 + C_2)},
\end{equation}
where $\mu_x$ and $\mu_y$ are the means of $x$ and $y$, $\sigma_x$ and $\sigma_y$ are the standard deviations of $x$ and $y$, $\sigma_{xy}$ is the covariance of $x$ and $y$, and $C_1$ and $C_2$ are constants. These metrics were calculated by comparing reconstructions with the ground truths for the mock data.

\begin{table}
	\centering
	\caption{Reconstruction quality for different values of $N$, where $N$ is the number of posterior 
	samples averaged over to create a reconstruction.}
	\begin{tabular}{|l|l|l|l|l|l|}
	\toprule
	N & PSNR $\uparrow$ & SSIM $\uparrow$ \\
	\midrule
	1 & 31.35 $\pm$ 0.01 & 0.6886 $\pm$ 0.0007 \\
	2 & 32.56 $\pm$ 0.01 & 0.7423 $\pm$ 0.0007 \\
	4 & 33.33 $\pm$ 0.01 & 0.7745 $\pm$ 0.0006 \\
	8 & 33.77 $\pm$ 0.01 & 0.7921 $\pm$ 0.0006 \\
	16 & 34.01 $\pm$ 0.01 & 0.8018 $\pm$ 0.0006 \\
	32 & \textbf{34.13 $\pm$ 0.01} & \textbf{0.8062 $\pm$ 0.0006} \\
	\bottomrule
	\end{tabular}
	\label{tab:results}
\end{table}

\begin{figure*}
	\centering
	\includegraphics[width=\textwidth, trim={0 1cm 0 0},clip]{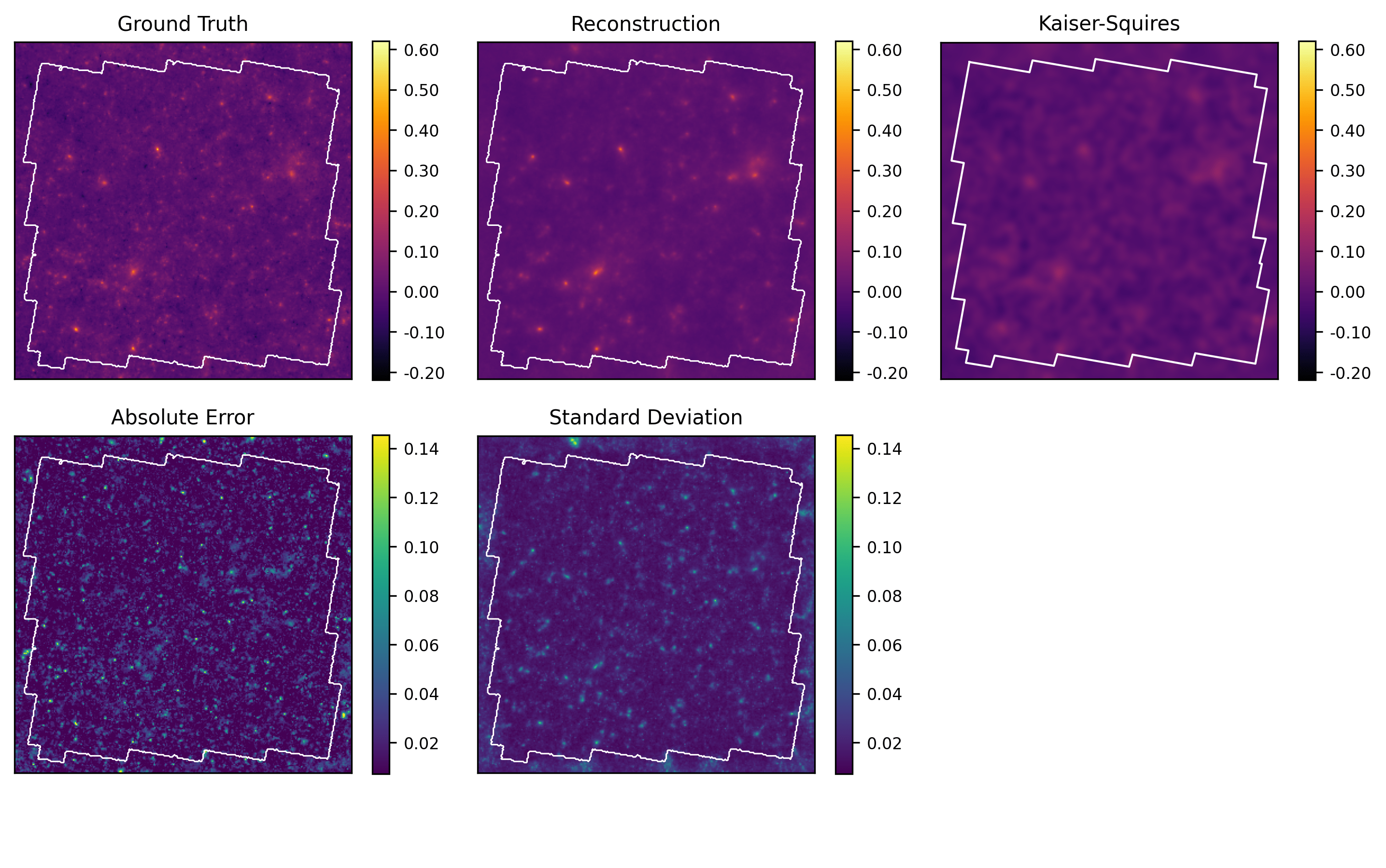}
	\caption{A reconstructed convergence map for one of the mock COSMOS maps.
	Our reconstruction is the average over 32 approximate posterior samples. On the bottom row is the pixel-wise absolute error between
	the reconstruction and the ground truth, and the standard deviation between the 32 samples 
	used to build the reconstruction. The white contour indicates the outer border of the mask applied to the data. We achieve superior visual quality as compared to the Kaiser-Squires reconstruction, with no peak suppression. Additionally, we see visual correlation between the absolute error and the standard deviation map.}
	\label{fig:overview_small}
\end{figure*}

Figure \ref{fig:overview_small} provides an overview of a given reconstruction, including the truth, a full reconstruction made by MMGAN, the absolute error between both, and the standard deviation of the approximate posterior samples used for the reconstruction. Note, the model was trained on masked data, so while it was able to fill small masked pixels within the central map, areas beyond the outer mask boundary (shown as a white contour on all figures) should be ignored, as the model was not trained there. We also show a Kaiser-Squires reconstruction, applying Gaussian smoothing here and throughout with variance $\sigma = 1 \, \text{arcmin}$, following \citet{remy2023probabilistic}, as this was shown to minimise the RMSE.

As can be seen in Figure \ref{fig:overview_small} our model has successfully captured the visual structure of the convergence map. The peaks are not suppressed in the reconstruction. The error between the truth and the reconstruction is very small in most areas. The same is true for the standard deviation. Importantly, from visual assessment, areas with the largest standard deviation correlate with areas of the highest error. This is sensical, as we expect that areas where the model is less certain of the true map, it will explore a wider range of possible reconstructions. Further examples for other simulated maps can be seen in the appendix, in Figure \ref{fig:big_overview}.

\begin{figure*}
	\centering
	\includegraphics[width=\textwidth]{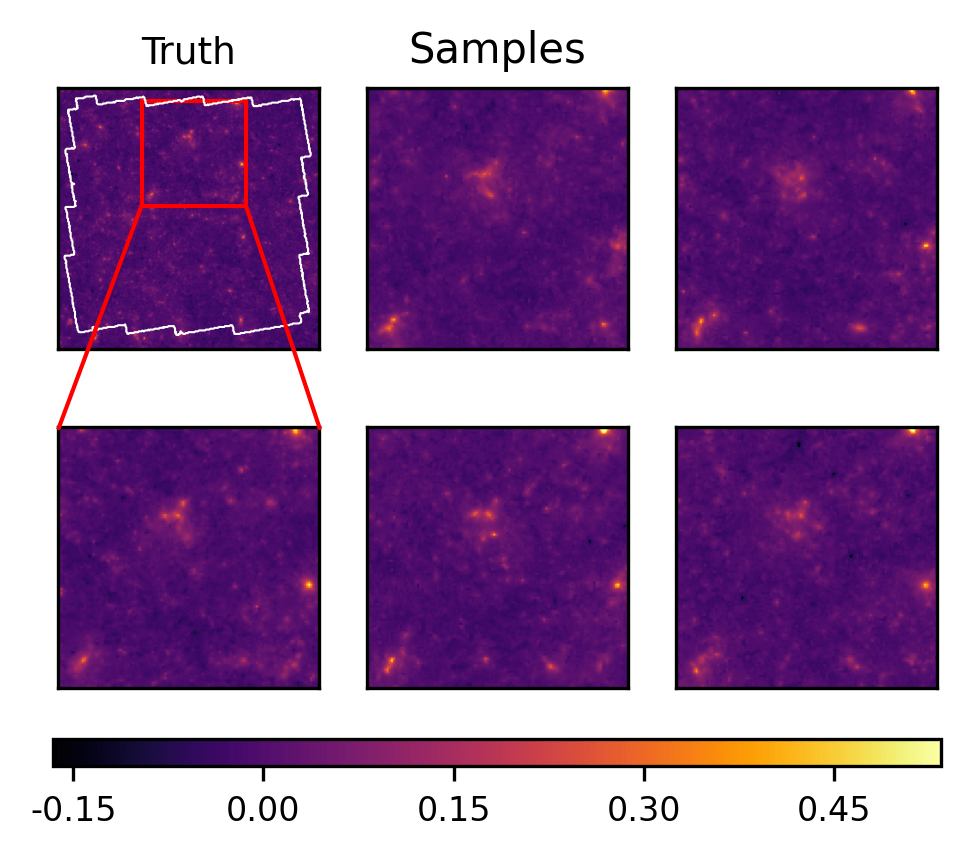}
	\caption{A selection of generated approximate posterior samples for a given shear map, in comparison 
	with the ground truth. We have zoomed in on a region of the samples, to better show the variation 
	within different samples.}
	\label{fig:samples}
\end{figure*}

As mentioned, MMGAN outputs samples from the learned posterior distribution. Figure \ref{fig:samples} shows a selection of generated samples, which highlights the sample generation diversity. Large scale features are consistent across the samples, however, the variability shows itself in the smaller scale structure, as can be seen in the differences in the zoomed-in regions of the figure.

\begin{figure*}
	\centering
	\includegraphics[width=\textwidth]{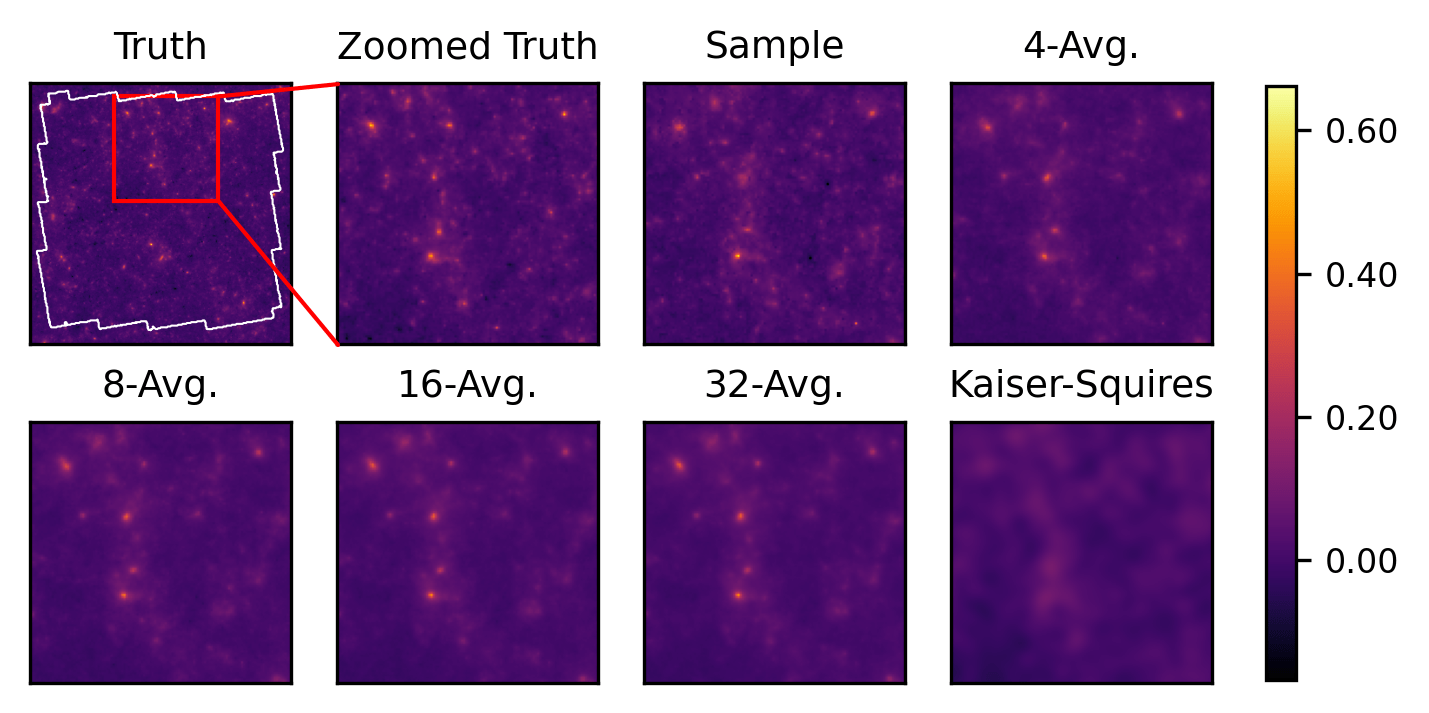}
	\caption{Demonstration of how the $N$-sample reconstruction varies 
	for $N \in \{1,4,8,32\}$, $N=1$ being a single posterior sample, for the zoomed-in region shown in the red box. The figure also shows the Kaiser-Squires map for the same 
	region. As can be seen, the reconstruction becomes smoother as $N$ increases, however, the 
	prominent features remain. An individual sample has a far higher level of detail, comparable with the true map, 
	however, it can be seen that features differ slightly to the truth, indicating why it is necessary to average over a number 
	of samples. Despite some loss of the smallest-scale structure for $N=32$, there is less peak suppression than the 
	Kaiser-Squires reconstruction.}
	\label{fig:zoomed_avg}
\end{figure*}

Figure \ref{fig:zoomed_avg} shows how the reconstruction varies as the number of samples used to build it changes. There is more detail when a 
smaller number of samples are used, however, these reconstructions are more prone to the variability of any individual posterior sample. By averaging 
over a larger number of samples, we do lose some level of small-scale structure, however, the features in the resulting reconstruction are more 
likely to be true features of the data. That said, even our 'smoother' reconstruction, with $N=32$, more accurately captures the small-scale structure to a higher level 
than the Kaiser-Squires reconstruction. Additionally, because we are not applying any additional post-processing---such as the Gaussian smoothing typical in Kaiser-Squires maps---there is no peak suppression of the small scale features.

\begin{table}
	\centering
	\caption{Results of validation metrics. The Pearson correlation coefficient, RMSE, and PSNR, were calculated for the Kaiser-Squires reconstruction (with $\sigma = 1 \, \text{arcmin}$ smoothing, chosen to minimise RMSE) and our 32-sample MMGAN reconstruction across a validation dataset. The results were averaged and then used to create this Table. Metrics for methods marked with an asterisk (*) are sourced from Table 1 in \citet{remy2023probabilistic} and therefore should not be directly compared with our results, since they consider a different validation set. Instead, they serve to provide a general comparison between MMGAN and other methods. Notably, the Kaiser-Squires results differ slightly from those reported in \citet{remy2023probabilistic}, likely due to differences in the randomly selected validation sets.}
	\begin{tabular}{|l|l|l|l|}
		\toprule
		  & Pearson $\uparrow$ & RMSE $\downarrow$ & PSNR $\uparrow$  \\
		\midrule
		MMGAN (Ours) & \textbf{0.727} & \textbf{0.0197} & \textbf{34.106} \\
		Kaiser-Squires & 0.619 & 0.0229 & 32.803 \\
		\midrule
		Kaiser-Squires * & 0.57 & 0.0240 & - \\
		Wiener filter * & 0.61 & 0.0231 & - \\
		GLIMPSE * & 0.42 & 0.0284 & - \\
		MCAlens * & 0.67 & 0.0219 & - \\
		DeepMass * & 0.68 & 0.0218 & - \\
		DLPosterior * & 0.68 & 0.0216 & - \\
		\bottomrule
	\end{tabular}
	\label{table:best_epoch}
\end{table}

In addition to visually comparing our MMGAN reconstructions to the Kaiser-Squires reconstruction, we also compare the two quantitatively through calculating a range of metrics. Those metrics are the Pearson correlation coefficient, RMSE, and PSNR. The results of this comparison can be seen in Table \ref{table:best_epoch}. MMGAN significantly outperforms Kaiser-Squires for each metric, which indicates that not only does our model produce reconstructions that visually appear to be of higher quality, but also that MMGAN is better capturing the underlying features of the data. Additionally, in Table \ref{table:best_epoch} we have included results from other state-of-the-art methods, as reported in \citet[\S Table 1]{remy2023probabilistic} including GLIMPSE \citep{lanusse2016high}, MCAlens \citep{starck2021weak}, DeepMass \citep{jeffrey2020deep}, and DLPosterior \citep{remy2023probabilistic}. These results are also using mock COSMOS data, built in the same way as our dataset. However, it is critical to stress that the validation set used to calculate these metrics differ, as can be seen from the difference in results for the Kaiser-Squires method. Therefore, the values in the table with an asterisk should not be compared directly with our own, however, they provide a general sense of MMGAN's performance with respect to other methods. In general, MMGAN, MCAlens, DeepMass, and DLPosterior all perform similarly well. Where MMGAN stands apart, is its ability to quantify uncertainties in a highly computationally efficient manner.

\subsubsection{Uncertainty Quantification Validation}\label{subsubsec:uq_validation}

As well as evaluating reconstruction quality, we also assessed the effectiveness of our uncertainty quantification. When building reconstructions of simulated convergence maps, we qualitatively compared the resulting uncertainty map with the pixel-wise absolute error between the MMGAN reconstruction and the ground truth. Visually, there is a correlation between these fields, as can be seen in Figure \ref{fig:overview_small} and Figure \ref{fig:big_overview}.

We also investigated MMGANs uncertainty maps using coverage tests, which assess whether the predicted uncertainties accurately reflect the true reconstruction error. Preliminary results suggest that MMGAN achieves strong performance. A comprehensive analysis of these tests will be presented in \citet{whitney2025}.

\subsection{COSMOS Field Reconstruction}\label{subsec:cosmos_results}

\begin{figure*}
	\centering
	\includegraphics[height=0.95\textheight]{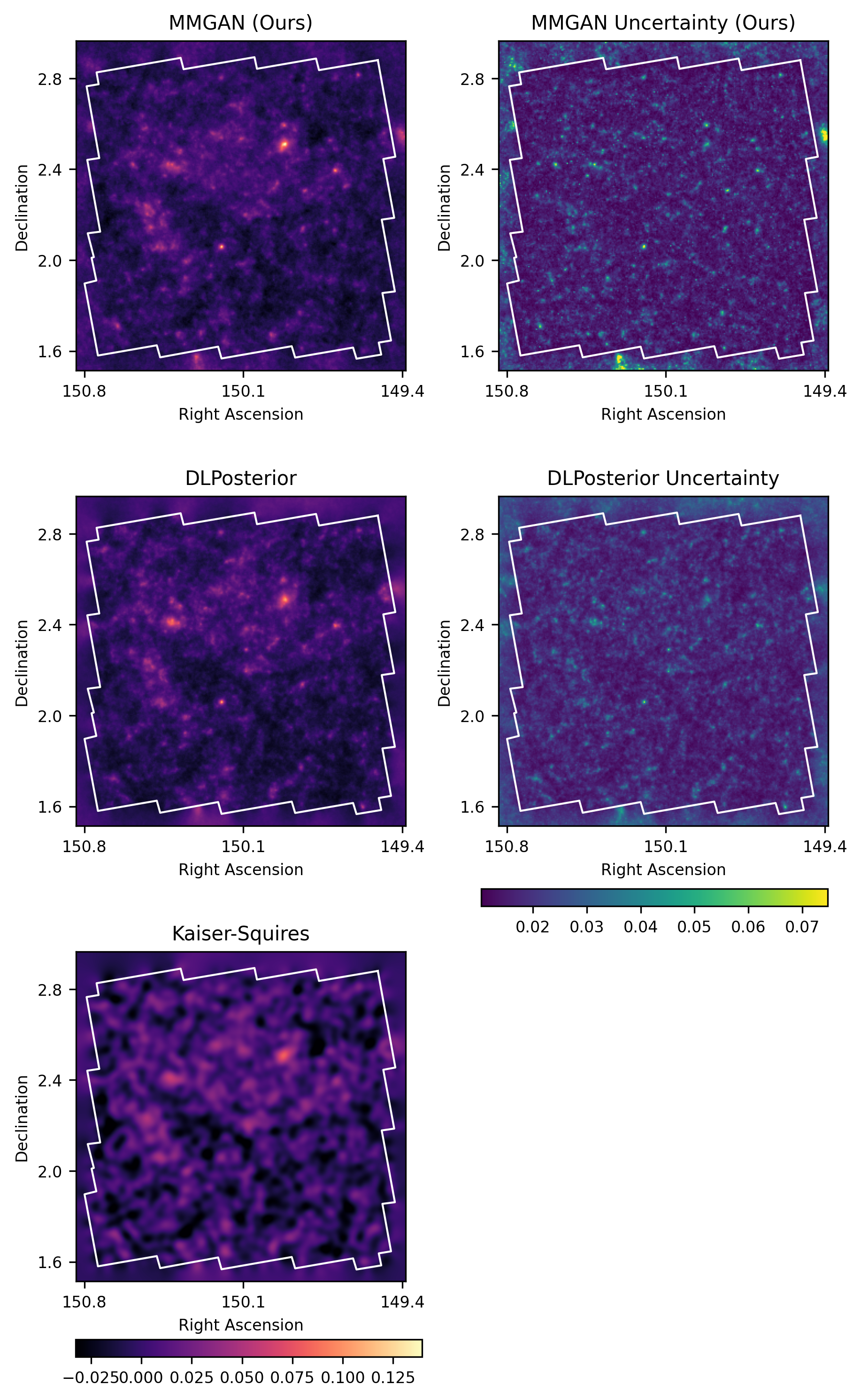}
	\caption{MMGAN reconstruction of the COSMOS field convergence map with uncertainties (\textit{top}), the DLPosterior reconstruction with uncertainties (\textit{middle}), and the Kaiser-Squires reconstruction (\textit{bottom}). All reconstructions are shown on the same scale; uncertainties also share a colour scale. The white contour indicates the outer border of the mask on the COSMOS field data.}
	\label{fig:cosmos_method_comp}
\end{figure*}

After validation we apply our full methodology to the COSMOS field data, using the catalog described in Subsection \ref{sec:cosmos_data}. Figure \ref{fig:cosmos_method_comp} shows an overview of our results. We compare with the DLPosterior COSMOS reconstruction of \citet{remy2023probabilistic}, in addition to the Kaiser-Squires reconstruction which acts as our baseline. Both our method and DLPosterior provide uncertainties, which are also included in Figure \ref{fig:cosmos_method_comp}. The three reconstructions are all shown on the same scale. 

When comparing the features present in our reconstruction with DLPosterior, we find good agreement in both the large and small scale structure. Peaks in the reconstructions are consistent in terms of magnitude and position.

Our reconstruction uncertainty is largely low throughout, with the highest magnitudes appearing in the masked region outside the COSMOS survey boundary. Again, the model was not trained to optimise this region, so a high level of uncertainty here is not surprising, and results in this region should be ignored. Interestingly, the uncertainties in the MMGAN reconstruction and the DLPosterior reconstruction are similar, with higher levels of uncertainty in the same regions. In order to better compare the uncertainties between MMGAN and DLPosterior reconstructions, we have shown them both on the same scale.

\begin{figure*}
	\centering
	\includegraphics[height=0.95\textheight]{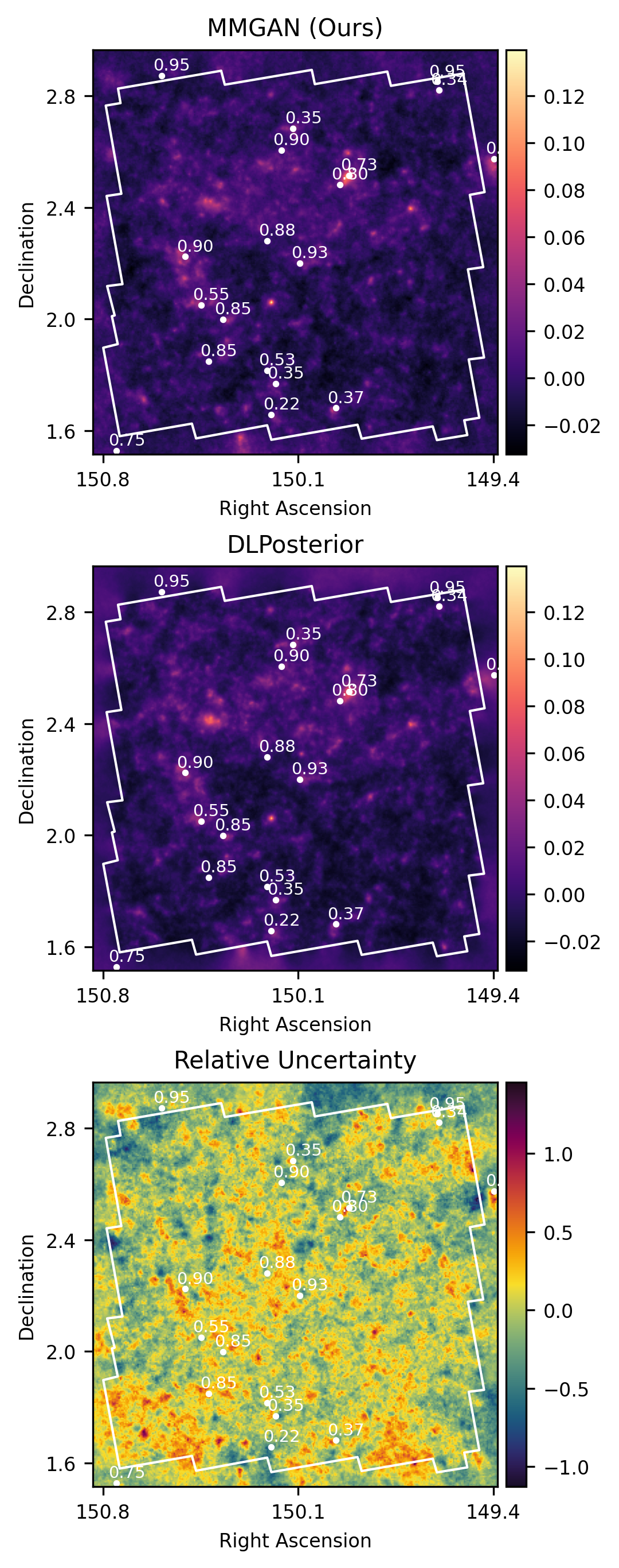}
	\caption{MMGAN COSMOS reconstruction (\textit{top}), the DLPosterior reconstruction (\textit{middle}), and the relative uncertainty between the two reconstructions (\textit{bottom}). The white points indicate the positions of known x-ray clusters from the \citet{finoguenov2007xmm} XMM-Newton data, and the white border is the edge of the COSMOS field mask. Both reconstructions are shown on the same scale. Both reconstructions are in good agreement with the x-ray data, and generally in good agreement with each other.}
	\label{fig:cosmos_rel_uq}
\end{figure*}

In order to draw a more detailed comparison between the reconstructions, we overlaid known x-ray clusters using a subset of the most massive clusters from the \citet{finoguenov2007xmm} XMM-Newton data, seen in Figure \ref{fig:cosmos_rel_uq}. We get good agreement between the features in our reconstruction and the cluster positions. There are a number of peaks in our reconstruction which do not have a corresponding cluster, however, given these features also appear in the DLPosterior and Kaiser-Squires reconstructions, they may be features which are beyond the depth of the x-ray data.

Another method of comparing our reconstruction with the DLPosterior reconstruction, is to take the relative uncertainty (RU) between the two using the following equation
\begin{equation}
	\text{RU} = \dfrac{M_1 - M_2}{\sqrt{S_1^2 + S_2^2}},
\end{equation}
where $M_1$ and $M_2$ are the convergence maps, and $S_1$ and $S_2$ are the standard deviations across the approximate posterior samples used for each reconstruction respectively. This equation can be interpreted as the number of standard deviations between the two reconstructions given the uncertainty estimated by each method. A low value means a high level of agreement between the two maps, and a high value indicates areas where the reconstructions do not agree as well. We show the relative uncertainty map between our reconstruction and the DLPosterior reconstruction  in Figure \ref{fig:cosmos_rel_uq}, as the lower panel. Overall the two reconstructions are in close agreement, with many pixels being within 1 standard deviation. The peaks in particular are in good agreement with one another, with the largest differences in the maps appearing in lower density regions in the reconstructions.

%=======================================================================================
\section{Conclusions}\label{sec:conclusions}
%=======================================================================================

Deep learning methods are a powerful tool in improving mass-mapping. They utilise data-driven priors, can handle the large amounts of data being collected by modern surveys, and are often better at capturing complex features in the data than traditional methods. However, in this era of precision cosmology, is it preferential that convergence map reconstructions which will be used for statistical analysis are accompanied by uncertainty maps. Most prior methods, including traditional and deep learning approaches, do not provide uncertainty estimates, and those that do can be slow. In order to address this gap, we propose MMGAN, a novel convergence map reconstruction method that provides uncertainties. MMGAN leverages a regularised conditional GAN to generate approximate posterior samples given shear observations, and then uses these samples to build a reconstruction and associated uncertainties. Under some assumptions, it can be shown that regularised conditional GANs are able to approximate the true posterior mean and standard deviation. Given these assumptions do not hold in all cases, an auto-tuning mechanism is adopted during training. 

Given a noisy shear observation, we construct a pseudo-reconstruction, and pass both into the MMGAN generator, which then outputs an approximate posterior sample. We take the posterior mean of $N=32$ approximate posterior samples as our final reconstruction, and the standard deviation of these samples to quantify our uncertainties. Currently, we choose a Kaiser-Squires map as our pseudo-reconstruction, however, this could be replaced with a more sophisticated reconstruction method, such as the Wiener filter. MMGAN does not require an explicit choice of cosmology, which is another reason why we chose the Kaiser-Squires method as our pseudo-reconstruction. We trained MMGAN on a mock dataset with a fixed cosmology, which we acknowledge may bias the learned model towards that cosmology. However, this is a limitation of the dataset rather than the method itself---there is no fundamental reason MMGAN could not be trained on a dataset with a range of cosmologies. Additionally, we did not account for astrophysical and observational systematics such as intrinsic alignments, photometric uncertainties, and shear multiplicative biases within our training data. These effects can introduce biases in the shear measurements, potentially impacting both reconstruction accuracy and uncertainty estimated. In future work, it would be interesting to explore how MMGAN performs when trained on data that accounts for these effects, as understanding their impact on the reconstruction and uncertainty maps is an important step towards ensuring the robustness of the method.

To train MMGAN, we used mock COSMOS-style shear and convergence maps, made from the $\kappa$TNG simulations. We validated our model on a subset of the mock COSMOS data not seen during training. We used the PSNR to ensure the model was suitably constrained in terms of variance in its sample generation, and then used standard metrics such as the PSNR, RMSE, and Pearson correlation coefficient to select the best training epoch of our model. We found that our MMGAN reconstructions are able to capture both large- and small-scale structure, and do not require any post-processing such as smoothing, which is known to suppress peaks. The resulting MMGAN model leverages data-driven priors to produce high-fidelity reconstructions with uncertainty estimates, all generated within seconds.

After validation we made a reconstruction of the COSMOS field, and found the results to be comparable to state-of-the-art methods, such as DLPosterior \citep{remy2023probabilistic}, and significantly more detailed than Kaiser-Squires. MMGAN was able to generate this reconstruction and associated uncertainties in under a minute, as compared to the $\sim$10 GPU minutes required to generate each independent approximate posterior sample by DLPosterior. The Kaiser-Squires reconstruction method \citep{kaiser1993mapping} and alternative deep learning approaches \citep{jeffrey2020deep,saxena2021generative}, while also being quick, provide no uncertainties.
Fast techniques that also quantify uncertainties are important for integration into downstream cosmological parameter estimation and model comparison pipelines so that uncertainties in the mass-mapping process are captured.

We hope our method will be useful in future mass-mapping analyses, in particular within larger pipelines that can make use of the rapid speeds at which posterior distribution samples can be generated. We make the code used for this work publicly available to the community, and hope it can be used to further the field of weak lensing mass-mapping.

%=======================================================================================
\section*{Acknowledgements}
%=======================================================================================

We thank Phillip Schniter and Matthew Bendel for insightful discussions. This work was supported by the Engineering and Physical Sciences Research Council (EPSRC) and the Science and Technology Facilities Council (STFC) through the following grants: EP/T517793/1; EP/W007673/1; ST/W001136/1. This work also used computing equipment funded by the Research Capital Investment Fund (RCIF) provided by UKRI, and partially funded by the UCL Cosmoparticle Initiative.
Figure \ref{fig:generator} was created using a tool developed by \citet{Iqbal2018PlotNeuralNet}.

%=======================================================================================
\section*{Data Availability}
%=======================================================================================

We provide the code used in this paper in a publicly available repository (\url{https://github.com/astro-informatics/rcGAN}), as well as the scripts used to reproduce the results. The repository is a fork of the original rcGAN repository by \citet{bendel2024regularized} (\url{https://github.com/matt-bendel/rcGAN}), on top of which substantial modifications and additions have been made.

We provide the trained model weights for the MMGAN model used in this paper, as well as the MMGAN COSMOS samples, reconstruction, and uncertainty maps in a publicly available Zenodo repository (\url{https://zenodo.org/records/14226221}).

%=======================================================================================
\bibliographystyle{mnras}
\bibliography{bibliography}

%=======================================================================================
\appendix

\section{Additional Simulation Plots}\label{sec:appendix}

\begin{figure*}
	\includegraphics[width=\textwidth]{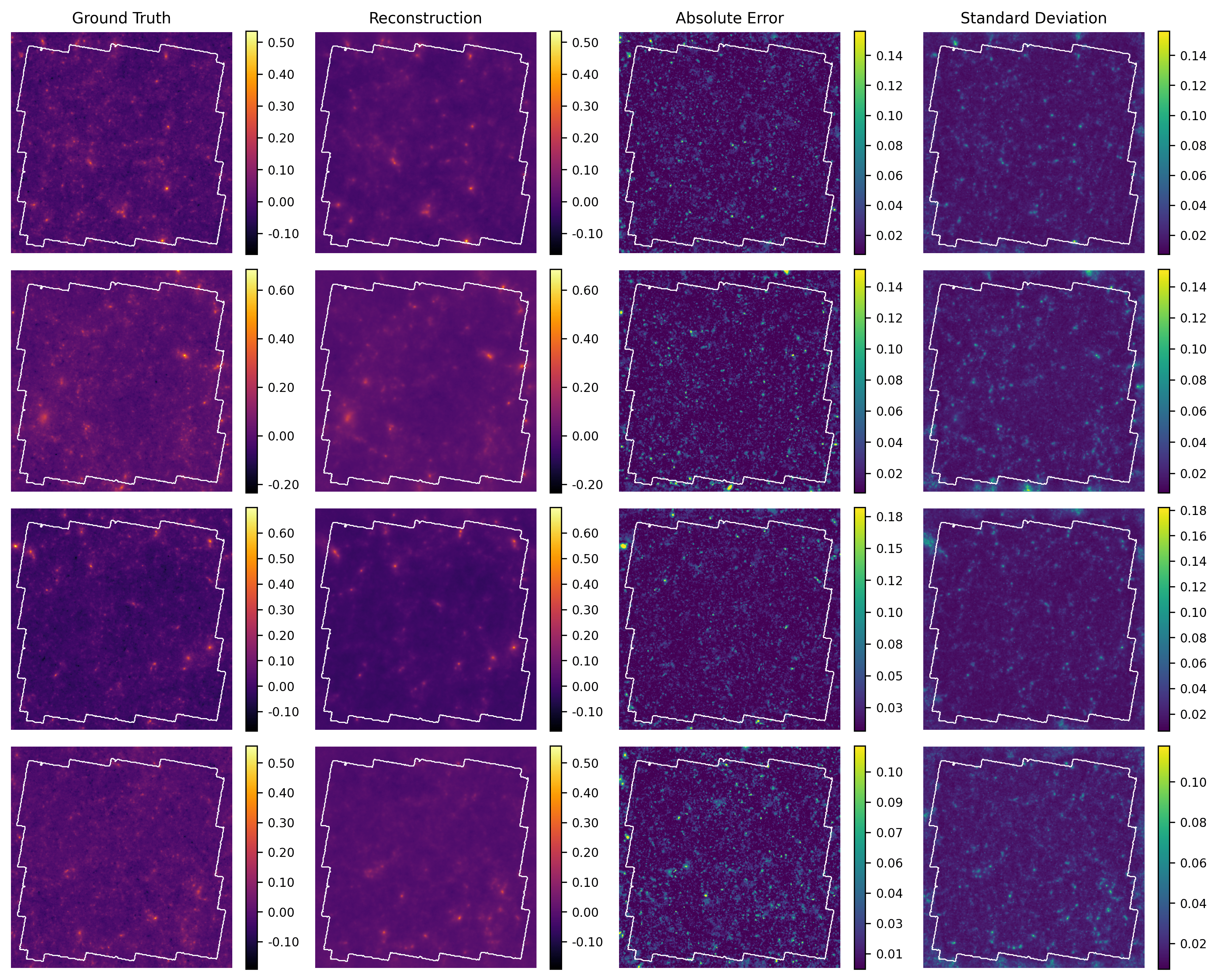}
	\caption{Overview of the MMGAN reconstructions of the COSMOS mocks including MMGAN reconstructions built from $N=32$ approximate posterior samples, the ground truth, the pixel-wise absolute error between the reconstruction and the ground truth, and the pixel-wise standard deviation between the 32 approximate posterior samples used to build the reconstructions.}
	\label{fig:big_overview}
\end{figure*}

In this section we provide an additional set of plots (\ref{fig:big_overview}) showing the MMGAN reconstructions of some simulated mock maps, alongside the ground truth, pixel-wise absolute error, and pixel-wise standard deviation. These plots are similar to those shown in Figure \ref{fig:overview_small}.

%=======================================================================================

% Don't change these lines
\bsp	% typesetting comment
\label{lastpage}
\end{document}